\documentclass[proof]{WileyASNA-v1CD}
\articletype{ORIGINAL ARTICLE}

\received{\today}
\revised{maybe}
\accepted{later}

\usepackage[rightcaption]{sidecap}
\usepackage{booktabs}
\usepackage{amsmath}
\usepackage{xcolor}
\usepackage{float}
\usepackage{alltt}

\usepackage{hyperref}
\hypersetup{
    final=true,
    pageanchor=true,
    colorlinks=true,
    breaklinks=true,
    linkcolor=blue,
    citecolor=blue,
    urlcolor=blue,
    pdfpagemode=UseNone,
    pdftitle={Solar observatory Einstein Tower - Data release of the 
        digitized solar full-disk photographic plate archive},
    pdfauthor={P. Pal et al.},
    pdfsubject={Solar Physics},
    pdfkeywords={sunspots, activity, photosphere, data analysis, 
        image processing}}

\usepackage[modulo, switch]{lineno}

\newcommand{\ns}{\hspace*{-5pt}}

\newcommand\arcdeg{\mbox{$^\circ$}} 
\newcommand\arcmin{\mbox{$^\prime$}}
\newcommand\arcsec{\mbox{$^{\prime\prime}$}} 

\begin{document}


\title{Solar observatory Einstein Tower --- Data release of the digitized
    solar full-disk photographic plate archive}

\author[1,2]{Partha S.\ Pal*}
\author[2]{M.\ Verma}
\author[2]{J.\ Rendtel}
\author[3]{S.J.\ Gonz{\'a}lez Manrique}
\author[2]{H.\ Enke}
\author[2]{C.\ Denker}

\authormark{Partha S.\ Pal \textsc{et al.}}

\address[1]{\orgname{Bhaskaracharya College of Applied Sciences,
University of Delhi}, \orgaddress{\state{Delhi}, \country{India}}}
\address[2]{\orgname{Leibniz-Institut f\"{u}r Astrophysik Potsdam (AIP)}, 
    \orgaddress{\state{Potsdam}, \country{Germany}}}
\address[3]{\orgname{Astronomical Institute of the Slovak Academy of Sciences}, 
    \orgaddress{\state{Tatransk{\'a} Lomnica}, \country{Slovak Republic}}}
    
\corres{*\email{ppal@aip.de}}

\presentaddress{Partha S.\ Pal,
    Leibniz-Institut f\"{u}r Astrophysik Potsdam (AIP),
    An der Sternwarte~16,
    14482 Potsdam,
    Germany}
    
\jnlcitation{\cname{%
    \author{P.S.\ Pal \textit{et al.}} (\cyear{2020}), 
    \ctitle{Solar Observatory Einstein Tower}, 
    \cjournal{ASNA}, \cvol{2020;00:1--8}.}}

\abstract[Abstract]{We present solar full-disk observations, which were recorded at the Einstein Tower during the years 1943\,--\,1991 (Solar Cycles 18\,--\,22). High-school students from Potsdam and Berlin digitized more than 3500 full-disk images during two- to three-week internships at AIP. The digital images cover a 15~cm $\times$ 15~cm region on photographic plates, which were scanned at a resolution of $7086 \times 7086$ pixels. The raw data are monochromatic 8-bit images in the Tagged Image File Format (TIFF). These images were calibrated and saved with improved photometric precision as 16-bit images with $2048 \times 2048$ pixels in the Flexible Image Transport System (FITS) format, which contains extensive headers describing the full-disk images and the observations. The various calibration steps include, for example, accurate measurements of the solar radius, determination of the limb-darkening function, and establishing an accurate coordinate system. The contrast-enhanced and limb-darkening corrected images as well as the raw data are freely available to researchers and the general public in publicly accessible repository. 
The data are published as a special data release of the Archives of Photographic PLates for Astronomical USE (APPLAUSE) project.}


\maketitle


\section{Introduction}

Regular scientific observations of the Sun and its activity started early in the $17^\mathrm{th}$ century with the invention of the first telescopes. Initially, sunspots and solar activity were recorded as daily drawings of the solar disk, and the recording techniques improved to provide nowadays full-disk images from space  with about one arcsecond spatial and a cadence of about 10~s. In the pre-photographic era, long time-series of hand-drawn full-disk images were collated by individual astronomers. Continuous efforts are underway to piece these records together, compiling a digitized archive of solar activity of the last 400~years. Full-disk solar images are collected with various setups: (1) filtergrams, where direct photographic imaging is performed with narrow-band filters to isolate features of strong chromospheric absorption lines, (2) spectroheliograms, where the full-disk images are obtained by scanning the solar disk using a spectrograph, and (3) images, where  broad-band filters are used for direct photographic imaging.

\citet{Arlt2008, Arlt2009a} digitized the drawings of Johann Caspar Staudacher covering observations of the years 1749\,--\,1796. This record was extended by \citet{Arlt2009b} to include drawings in the period 1795\,--\,1797 by James Archibald Hamilton and William Gimingham. Further work of e.g., \citet{Arlt2013} and \citet{Karoff2019} covered observations taken by Samuel Heinrich Schwabe in 1825\,--\,1867, while \citet{Diercke2015} digitized drawings of Gustav Sp\"orer for the period 1861\,--\,1894. This subjective list of digitization projects is by no means exhaustive. Many past and ongoing efforts are aimed at preserving and making accessible a digitized record of historical solar observations. The aforementioned datasets of digitized drawings were used for a detailed analysis of the sunspot properties. For example, \citet{SenthamizhPavai2016} employed a time-series covering about 270~years for studying the sunspot group tilt angle. The authors concluded that the average tilt angles before the Maunder Minimum were similar to modern values, however, the average tilt angles of a period 50 years after the Maunder minimum, i.e., for Cycles~0 and~1, were much lower and near zero.

Regular photographic observations with the Kew and Dallmeyer photoheliographs of the Sun started in 1874 at the Royal Observatory, Greenwich, United Kingdom, which were soon adapted at other solar observatories around the world \citep{Baumann2005, Willis2016}. Modern computer technology and advanced image processing software facilitated the creation of digital archives in the last decade, preserving the photographic record of solar full-disk observations and making them available for contemporary research. Several projects were undertaken to digitize historical data obtained at many solar observatories including Kodaikanal Solar Observatory (India), Arcetri Astrophysical Observatory (Florence, Italy), Astronomical Observatory of Coimbra (Portugal), and Mt.\ Wilson Solar Observatory (California), among others.

Kodaikanal Solar Observatory acquired 31,800 plates covering over 31,000 days of observation since its inception in 1904. The white-light telescope consists of a 10~cm objective lens with an $f/15$ beam that is capable of producing a 20.4-cm diameter solar image. In 1918, the optics was upgraded to a 15-cm achromatic lens with focal length of 240~cm. Details of the digitization of this 100-year data volume are presented by \citet{Ravindra2013}, who also describe the identification of sunspots in digitized images, the estimation of sunspot area and sunspot group membership on each day, and the detailed comparison with the data of the Royal Observatory, Greenwich.

In 1926\,--\,1974, Arcetri Astrophysical Observatory recorded photographic plates of solar full-disk spectroheliograms, covering the cores of the strong chromospheric absorption lines Ca\,\textsc{ii}\,K and H$\alpha$ \citep{Ermolli2009a, Ermolli2009b}. In total, 12,917 plates are available in the archive, comprising 5976 Ca\,\textsc{ii}\,K and 6941 and H$\alpha$ spectroheliograms, respectively. The grating of the spectrograph had a size 100~mm $\times$ 100~mm with 600 lines mm$^{-1}$ and a dispersion of 0.33~mm/{\AA} at 3934~\AA. \citet{Ermolli2009b} also conducted preliminary measurements of position and area of facular regions, which were identified in the digitized Ca\,\textsc{ii}\,H spectroheliograms.

Astronomical Observatory of Coimbra (COI) acquired solar images since 1925 and maintains one of the largest continuous solar data collections in the world. The optical system of the spectroheliograph is composed of a coelostat and a horizontal telescope, where the objective lens has a focal length of $f = 406$~cm and diameter of $D =25$~cm. The coelostat is composed of two mirrors with diameters of 40~cm and 30.5~cm, respectively. \citet{Carra2018} catalogued the historical sunspots observation for the period 1921\,--\,1941 and estimated their extent in longitude and latitude limits as well as the umbral area of sunspots. The catalogue of full-disk data comprises 20,505 Ca\,\textsc{ii}\,K line-core images, 11,612 images of the minima in the emission core of the Ca\,\textsc{ii}\,K line, 8719 H$\alpha$ line-core images accompanied by 3889 nearby continuum images, and 3024 dopplergrams \citep{Louren2019}. More recently, a detailed study of the penumbra-umbra ratio was conducted by \citet{Carrasco2018} using these data.

The Mt.\ Wilson Solar Observatory was founded in 1904. The 60-foot Solar Tower was completed in 1908. The vertical telescope has a 61-cm primary mirror with a focal length of $D = 18$~m feeding a spectrograph, which is in regular use since 1915. Over 150,000 images of the Sun were acquired in the last 100 years. The images include more than 43,000 white-light image as well as more than 35,000 Ca\,\textsc{ii}\,K and 74,000 H$\alpha$ spectroheliograms \citet{Lefe2005}. \citet{Bert2010} introduced a new Ca\,\textsc{ii}\,K plage and active network index time-series derived from the digitization of almost 40,000 photographic plates that were obtained in the period 1915\,--\,1985. \citet{Bert2008} estimated the rate of solar rotation over the whole solar surface as a function of time using the day-to day motions of contrast features in these  Ca\,\textsc{ii}\,K spectroheliograms for the period 1915\,--\,1975. In \citet{Tlatov2009}, a new method for digitizing and calibrating Ca\,\textsc{ii}\,K photographic plates is introduced for data holdings at Kodaikanal Observatory, Mt.\ Wilson Observatory, and National Solar Observatory at Sacramento Peak.

Full-disk solar imaging evolved significantly from hand-drawings via photographic plates to digitized images from large-format CCD detectors with good spatial and temporal resolution as well as with accurate photometry. In addition, while ground-based full-disk observations were historically carried out at observatories at or in proximity to astronomical institutes, nowadays full-disk telescopes are located at the best observing sites and often form networks of identical telescopes, spread across to globe in longitude, to continuously follow the evolution of solar features and to monitor long-term changes of solar activity.

For example, the Global Oscillation Network Group \citep[GONG,][]{Harvey1996} funded by the U.S.\ National Science Foundation (NSF) studies the internal structure and dynamics of the Sun. The network started gathering continuous solar data in 1995 from six identical instruments distributed around the world. The robotic instruments measure Doppler shifts of the photospheric Ni\,\textsc{i} $\lambda$676.8~nm line, which are used to derive the $p$-modes of the 5-minute oscillation. Continuum and line-of-sight (LOS) magnetograms are also available since 1995. The network was upgraded in 2000 to include H$\alpha$ full-disk filtergrams, complementing data of the global high-resolution H$\alpha$ Network \citep{Steinegger2000b} with Big Bear Solar Observatory (BBSO) in California,  Kanzelh\"{o}he Solar Observatory in Austria, and Yunnan Astronomical Observatory in China as founding members. 

The Chromospheric Telescope \citep[Chrotel,][]{Kentischer2008, Bethge2011} is another robotic telescope installed at the Observatorio del Teide, Tenerife, Spain, which acquires full-disk filtergrams of the Sun in the chromospheric Ca\,\textsc{ii}\,K, H$\alpha$, and He,\textsc{i} $\lambda$10830\AA\ spectral lines. The Solar and Heliospheric Observatory \citep[SoHO,][]{Domingo1995}, represents the last leap in synoptic full-disk observations of the Sun, entering the era of space-missions with long-term programs furnishing the high-cadence images as well as magnetograms. The very successful SoHO mission was superseded by the Solar Dynamics Observatory \citep[SDO,][]{Pesnell2012}, which started operating in 2010. Besides continuum images and magnetograms, SDO provides high-cadence full-disk E/UV images with good spatial resolution capturing the dynamics of atmospheric layers from the photosphere to the corona. This list of synoptic solar observations is certainly incomplete. A more comprehensive summary is given in \citet{Pevtsov2016}.

In the following, full-disk observations of the solar observatory Einstein Tower are described, which includes a detailed account of the observations, data reduction, and the various levels of data products. The digitized photographic plates will become part of the Archives of Photographic PLates for Astronomical USE (APPLAUSE) project.\footnote{\href{https://www.plate-archive.org/applause/documentation/dr3s/}{www.plate-archive.org/applause/documentation/dr3s/}}


\section{Solar Observatory Einstein Tower}

\begin{figure}[t]
\centerline{\includegraphics[width=0.5\textwidth]{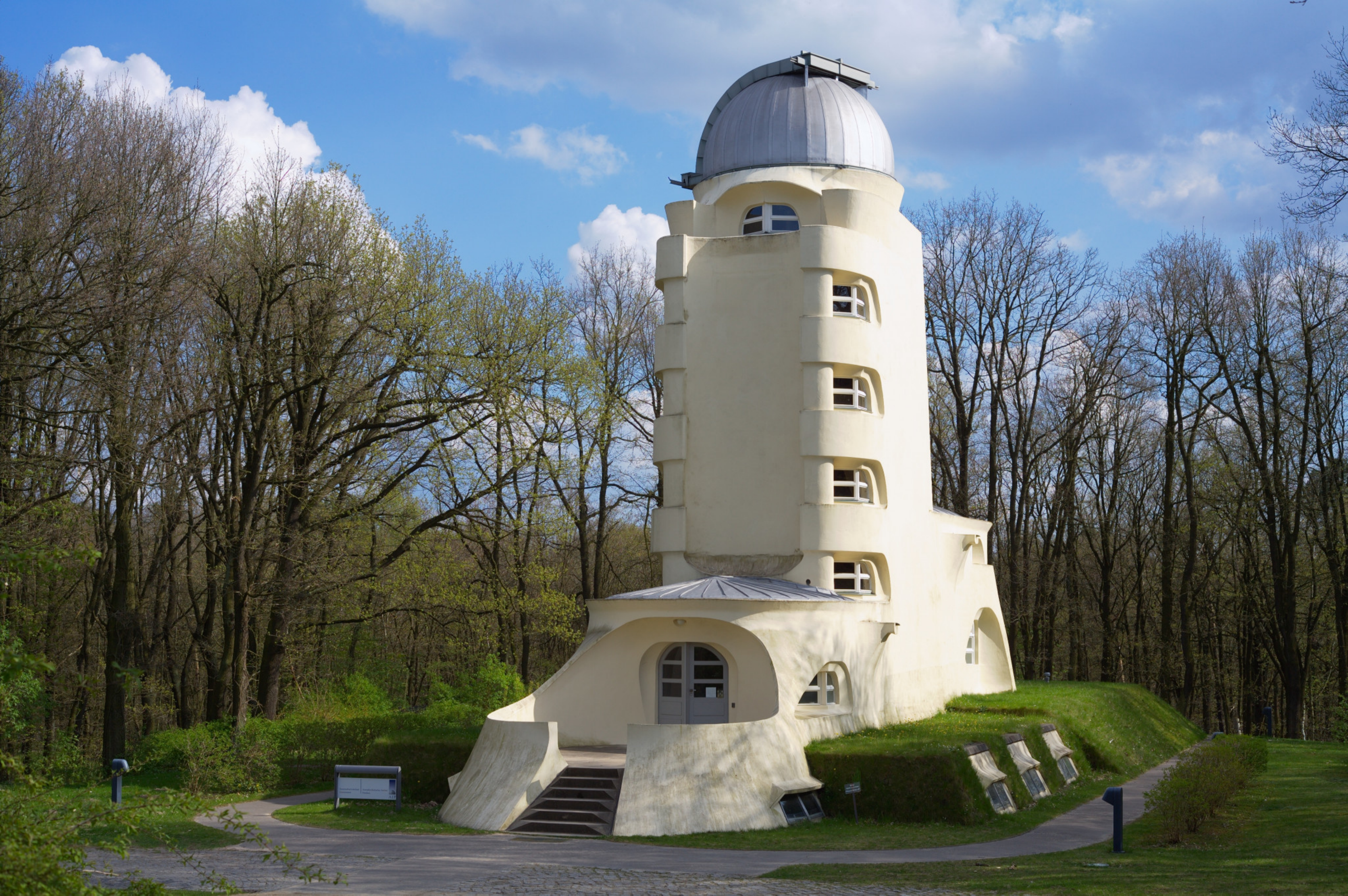}}
\caption{The Einstein Tower, located within the Albert-Einstein Science Campus 
    at the Telegraphenberg in Potsdam, is part of the Leibniz-Institut f\"ur
    Astrophysik Potsdam (AIP).}
\label{FIG01}
\end{figure}

The Einstein Tower (Fig.~\ref{FIG01}\ns) is the first important building designed by the architect Erich Mendelsohn exemplifying architectural expressionism \citep{Denker2016}. It was planned and built in the years 1919\,--\,1924, with the main part already finished in 1921. The Einstein Tower was conceived as an instrument to confirm the gravitational redshift of solar spectral lines predicted by general relativity. Substantially overhauled since November 1997, the Einstein Tower re-opened on 1999 July~1. The observatory is still a functional building with operational instruments, which nowadays are mainly used for instrument development and calibration before deployment to observing facilities such as the GREGOR solar telescope on Tenerife \citep{Schmidt2012} or the Large Binocular Telescope in Arizona \citep{Hill2006a, Strassmeier2015}. A detailed description of the architectural achievements and the historical background are presented in the monographs by \citet{Hentschel1992} and \citet{Wilderotter2005}.

Build as the first solar tower telescope in Europe, two 85-cm aperture coelostat mirrors at a height of 15~m above ground direct the light vertically into the tower. There the sunlight encounters a 60-cm aperture objective lens with a focal length of $f = 14$~m, which produces a solar image with a diameter of about 13~cm in the optical laboratory located in the basement of the building. The telescope is supported on a wooden construction with its own foundation to avoid wind buffeting and other vibrations. The basement location of the optical laboratory ensures a stable thermal environment, which is required for high resolution spectroscopy. A plane fold mirror before the prime focus directs the light horizontally to a high resolution spectrograph (main spectrograph, fourth order: ${\cal R}_\mathrm{max} = \lambda / \Delta\lambda = 1,000,000$ at $\lambda$630~nm), which has its own room separated from the main observing room \cite{Staude1991, Staude1995}.

\begin{figure}[t]
\centerline{\includegraphics[width=0.5\textwidth]{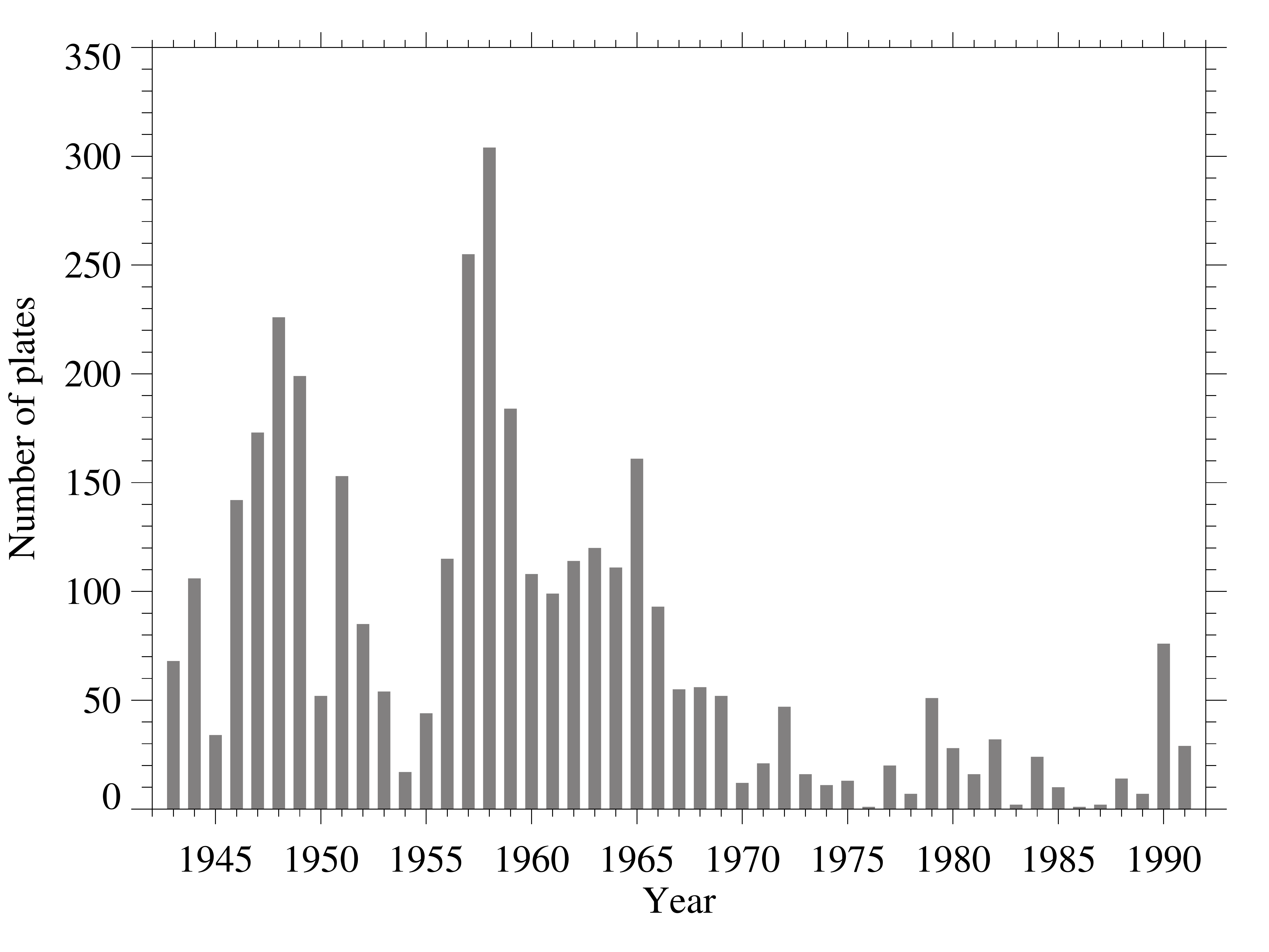}}
\caption{Data coverage of photographic plates in the years 1943\,--\,1991. Only sporadic observations were carried out after 1970.}
\label{FIG02}
\end{figure}

Preserving scientific heritage and making it usable for contemporary scientific exploitation is the primary goal of the APPLAUSE project. In solar physics, the historical data are required to fill gaps in decade- and even century-long time-series to improve our understanding of solar activity variations and the 11-year activity/22-year magnetic cycle of the Sun. As part of the APPLAUSE project, more than 3500 full-disk images were digitized by high-school students from Potsdam and Berlin during two- and three-week internships at Leibniz-Institut f\"ur Astrophysik Potsdam (AIP). The digitized record of solar photographic plates comprises the years 1943\,--\,1990. The data coverage of the daily full-disk images per year and as a function of observing time are displayed in Figs.~\ref{FIG02}{\ns} and~\ref{FIG03}\ns, respectively. Their calibration and digitization are described in the following sections including an introduction of the final data products. 

\section{Observations}

The geographic location of the solar observatory Einstein Tower is \mbox{52\arcdeg~22\arcmin~44\arcsec~N} and \mbox{13\arcdeg~3\arcmin~49\arcsec~E}, where it is located on Potsdam's Telegraphenberg at an altitude of about 90~m. Solar images with a diameter of 13~cm were captured on photographic plates and in some cases on photographic film since 1943. In general, the quality of the photographic material is high. However, insufficient flushing after chemical development leads to some deterioration of the photographic material. In most cases, the residual haze can be carefully wiped off with microfleece without leaving any trace. Some newer images contain a calibration step wedge, which can be potentially used for densitometry and photometric corrections. However, this information was not used so far because uniform processing of all images took precedence.

The camera set-up, operation, and observing procedure is presented in \citet{vonKlueber1948}. Figure~\ref{FIG04}{\ns} shows the schematic of the slit camera used at the Einstein Tower to obtain photographic plates is taken from \citet{vonKlueber1948}. The various parts for this slit-camera are labeled $a$\,--\,$t$. The schematic and working of the camera is as follows: ($a$) polished rail of round-bar steel works as the guide rail for the shutter with a movable rectangular cover plate ($b$), which is mounted on ball bearing rolls ($c$ \& $d$) ensuring a uniform exposure. The shutter plate has a vertical slit ($e$), which can be precisely adjusted in width using an adjustable slit-jaw ($f$) and knurled-head screw ($g$). The adjustment of the slit-width is measured by a scale ($h$). The slit-jaw is clamped on the shutter plate using tommy screws ($i$ \& $j$). The large-format photographic plates ($k$) are mounted in a wooden frame and a spring-loaded mechanism ($l$\,--\,$q$) is utilized to release the cover plate for an exposure. After the first exposure, a additional system ($r$\,--\,$t$) is used to put directional marks and an intensity scale via a second exposure of the plate. Further details for capturing of photographic plates are presented in \citet{vonKlueber1948} and \citet{Denker2016}.

\begin{figure}[t]
\centerline{\includegraphics[width=0.5\textwidth]{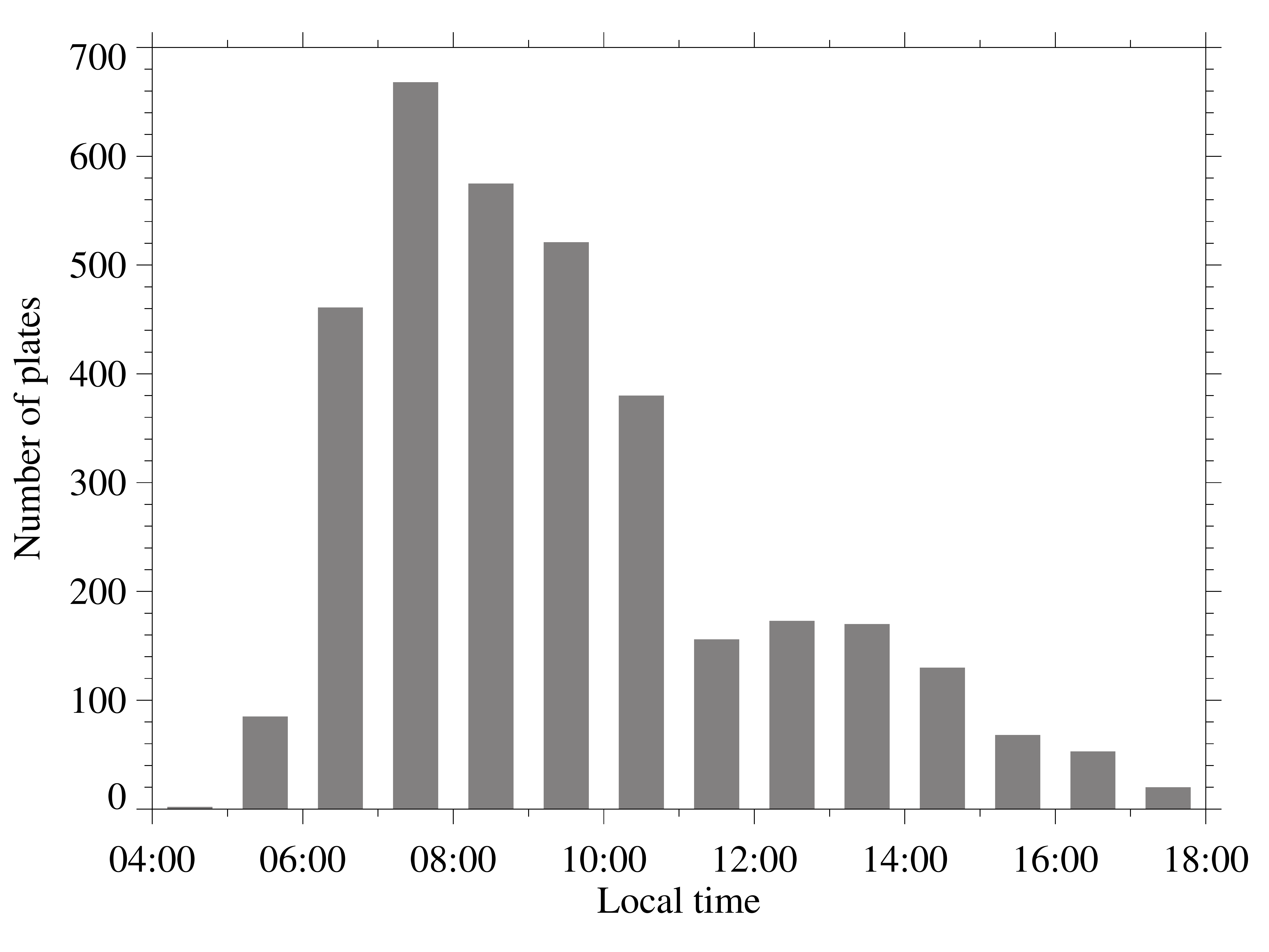}}
\caption{Data coverage of photographic plates during the observing day.}
\label{FIG03}
\end{figure}

More than 3500 plates were recorded from 1943\,--\,1991. The regular full-disk observations using photographic plates and film stopped after that. Figure~\ref{FIG02}{\ns} shows the number of observations recorded for almost half a century. The first half of the dataset comprises most of the images, i.e., around 128 images per year. This number dropped to around 22 images per year in the second half. In addition, most of the plates are captured in the first few hours after sunrise, i.e., in the time period 07:00\,--\,10:30~CET as evident in Fig.~\ref{FIG03}\ns. Some plates were taken as early as 05:30~CET. The latest time used for observation was 17:30~CET. Early morning observations are typically a good compromise between decreasing air mass and increased ground heating as the Sun rises in elevation. The image quality is in general very good, i.e., umbrae and penumbrae of sunspots are normally easily separable -- in exceptional cases even sunspot fine structure such as penumbral grains and umbral dots are visible.

\begin{figure}[t]
\centerline{\includegraphics[width=0.5\textwidth]{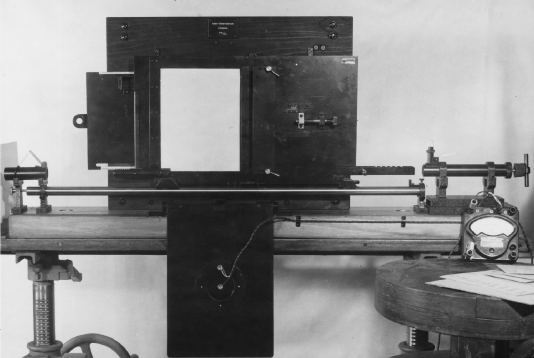}}
\centerline{\includegraphics[width=0.5\textwidth]{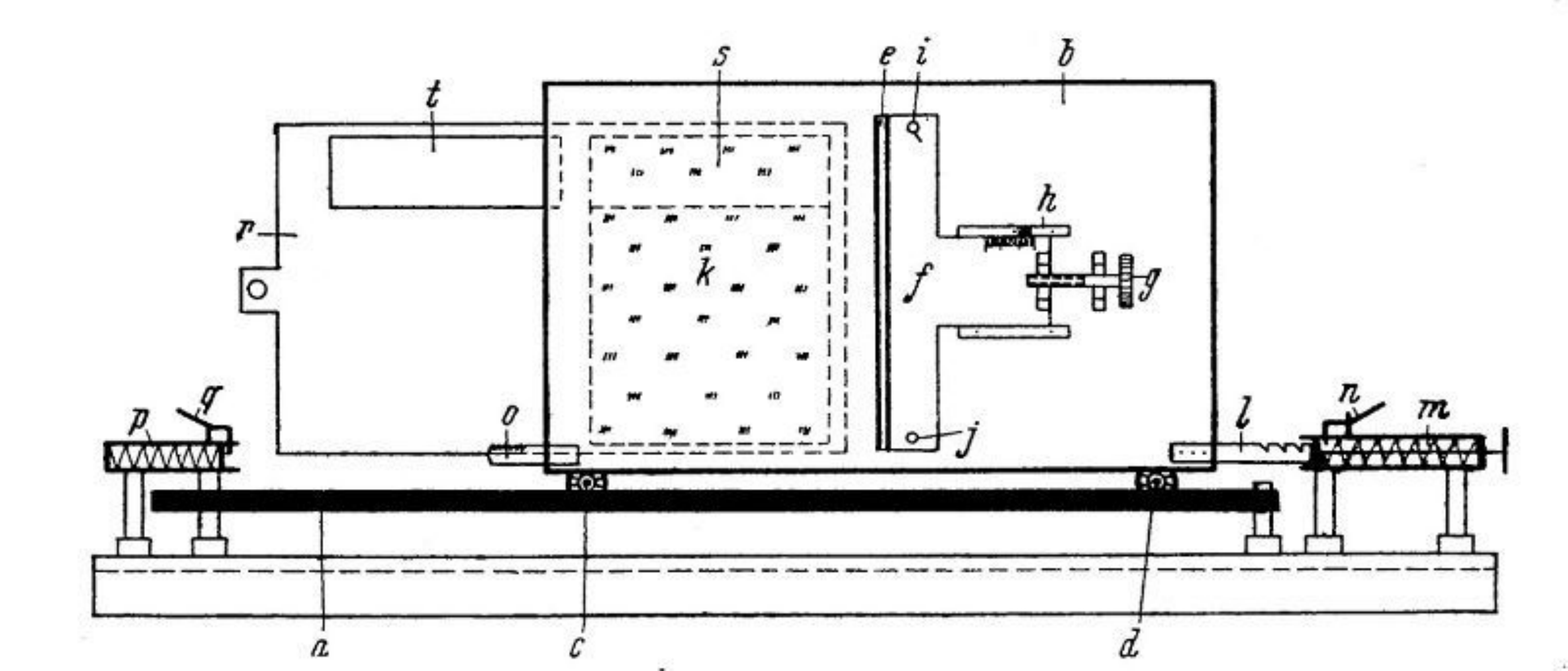}}
\caption{Photograph of the slit camera at the Einstein Tower that was used for recording solar full-disk images (\textit{top}). Schematic sketch of the slit camera (\textit{bottom}). The parts of the camera are labeled $a$\,--\,$t$ and explained in the text \citep[adapted from][]{vonKlueber1948}. More details are presented in \citet{Denker2016}.}
\label{FIG04}
\end{figure}

\begin{figure}[t]
\centerline{\includegraphics[width=0.5\textwidth]{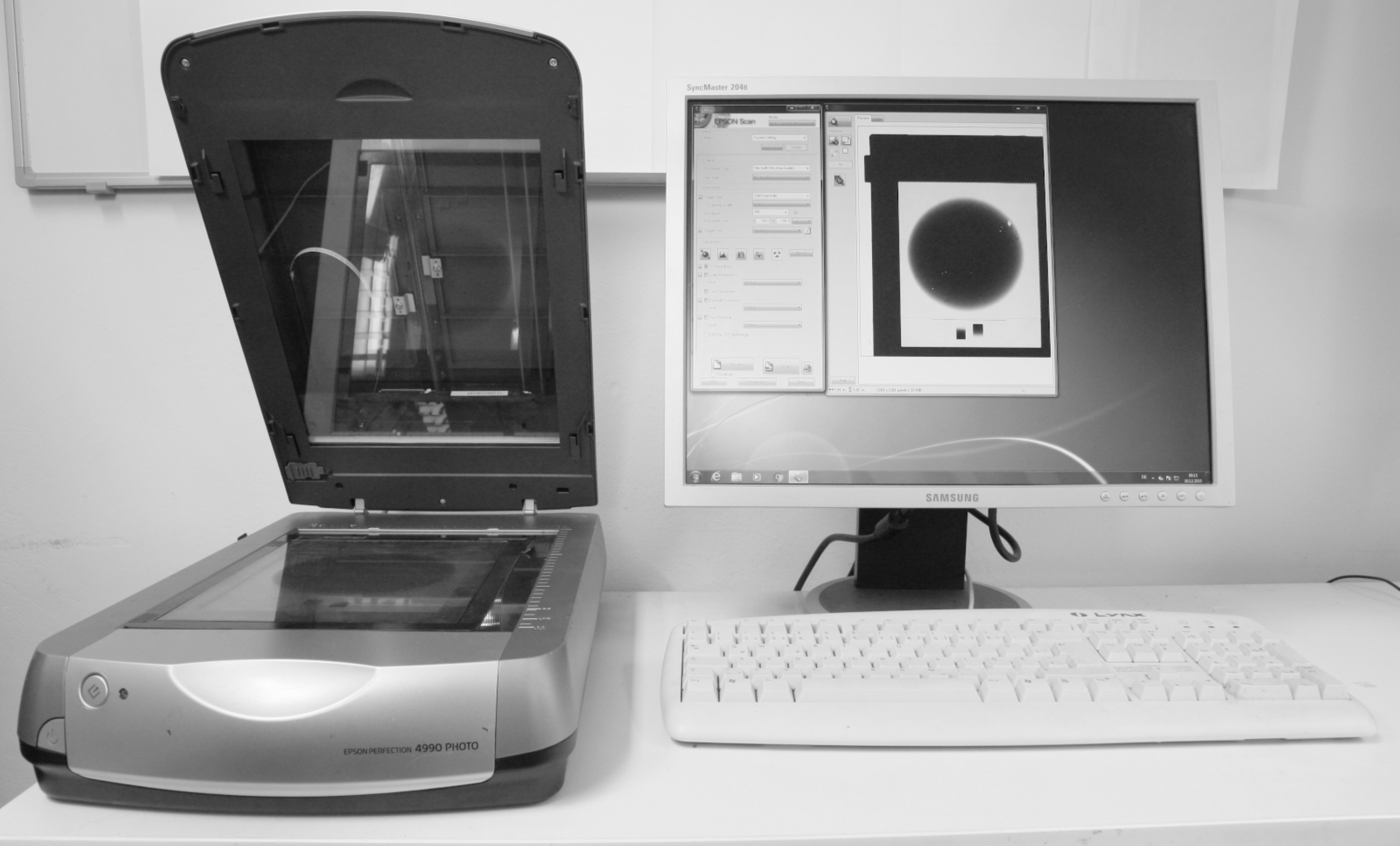}}
\caption{Epson flatbed scanner (\textit{left}) that was used for the digitization of the photographic plates. The monitor (\textit{right}) displays the graphical user interface of the scanning software with a display window containing the metal mask, which encompasses a carefully aligned full-disk image.}
\label{FIG05}
\end{figure}

The archived photographic plates are scanned with an ``Epson Perfection~4990 Photo'' flatbed color scanner with USB~2.0 interface. The photoelectric device of the scanner is a CCD detector, which provides an optical resolution of up to $4800 \times 9600$~dpi (horizontal $\times$ vertical) for a scan area of 216~mm $\times$ 297~mm. However, the resolution was set to 1200~dpi in both directions. The light source is a white cold cathode fluorescent lamp. Only a region of 15~cm $\times$ 15~cm ($7086 \times 7086$ pixels) on the photographic plates is scanned and saved as monochromatic 8-bit image in the Tagged Image File Format (TIFF). The precise selection of the scan region is achieved with a metal mask matching exactly the dimensions of the photographic plates. Scanner and mask are displayed in Fig.~\ref{FIG05}\ns, which also shows the graphical user interface of the scanning software, Sample datasets were already presented in \citet{Denker2016}.

  
\section{Data Reduction}

The goal of the data reduction is to carefully calibrate the scanned full-disk images and to save these data with improved photometry as $2048 \times 2048$ pixel images in the Flexible Image Transport System \citep[FITS,][]{Wells1981, Hanisch2001} format. This format allows us to associate metadata with the images as standardized ASCII headers, which is a prerequisite in making the data accessible for contemporary scientific use. Ultimately, contrast-enhanced and limb-darkening corrected images will be publicly available to researchers and the general public within the framework of the APPLAUSE project. In addition, the metadata will be stored in a relational database, which allows users custom queries and in the future the addition of feature catalogues, containing for examples sunspot properties extracted from the full-disk images.

\begin{figure*}[ht]
\centerline{\includegraphics[width=0.95\textwidth]{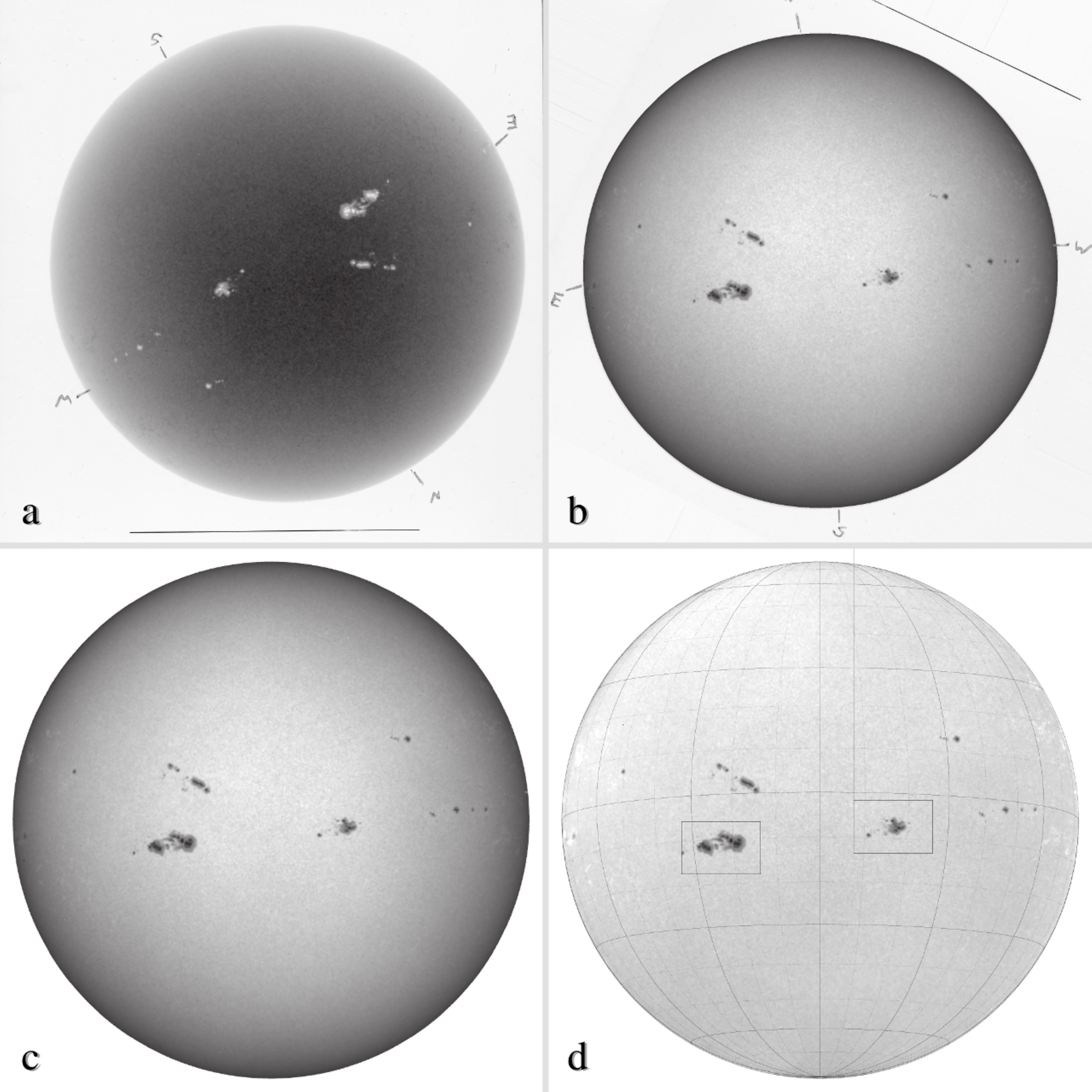}}
\caption{(a) Photographic negative of a broad-band full-disk image obtained at the solar observatory Einstein Tower on 1949 February~3. The solar orientation was marked directly by hand on the photographic material. The stripe at the bottom results from a pinhole on the moving slit shutter indicating the scanning direction. (b) Photographic positive with standard solar orientation (North up and East left) after correction of the position angle $P = -13.2^\circ$. (c) Calibrated and centered image (Level~1) with removed background artefacts. (d) Calibrated image with limb-darkening correction (Level~2). Heliographic coordinates are indicated at intervals of $30^\circ$ (\textit{solid lines}) and $10^\circ$ (\textit{dotted lines}), respectively. Enlarged versions of the two rectangular boxes are presented in Fig.~\ref{FIG07}\ns.}
\label{FIG06}
\end{figure*}

The data reduction pipeline is very similar to the one for H$\alpha$ full-disk observations at Big Bear Solar Observatory \citep{Denker1999a}, however, with modifications adapted to the digitized photographic broad-band images. The full-resolution (7k $\times$ 7k pixels), full-disk images are digitized as 8-bit integers. The photographic negative images (see Fig.~\ref{FIG06}{\ns}a) are simply inverted in the first step (Level~0 data) and saved as 8-bit images in the FITS format with basic information stored in the ASCII header. Sample FITS headers for the various data levels are compiled in Sect.~\ref{LEVEL}. The original TIFF data are not part of the digital archive. The data processing is carried out on standard personal computers using custom programs in the Interactive Data Language\footnote{\href{https://www.harrisgeospatial.com/}{www.harrisgeospatial.com}} (IDL), which will become part of the sTools software library \citep{Kuckein2017}. In addition, general purpose routines of SolarSoft\footnote{\href{https://www.lmsal.com/solarsoft/}{www.lmsal.com/solarsoft/}} \citep{Bentley1998, Freeland1998} and  NASA's IDL Astronomy User's Library\footnote{\href{https://idlastro.gsfc.nasa.gov/}{idlastro.gsfc.nasa.gov/}} are used mainly for handling of the FITS data.

\begin{figure*}[t]
\centerline{\includegraphics[width=0.49\textwidth]{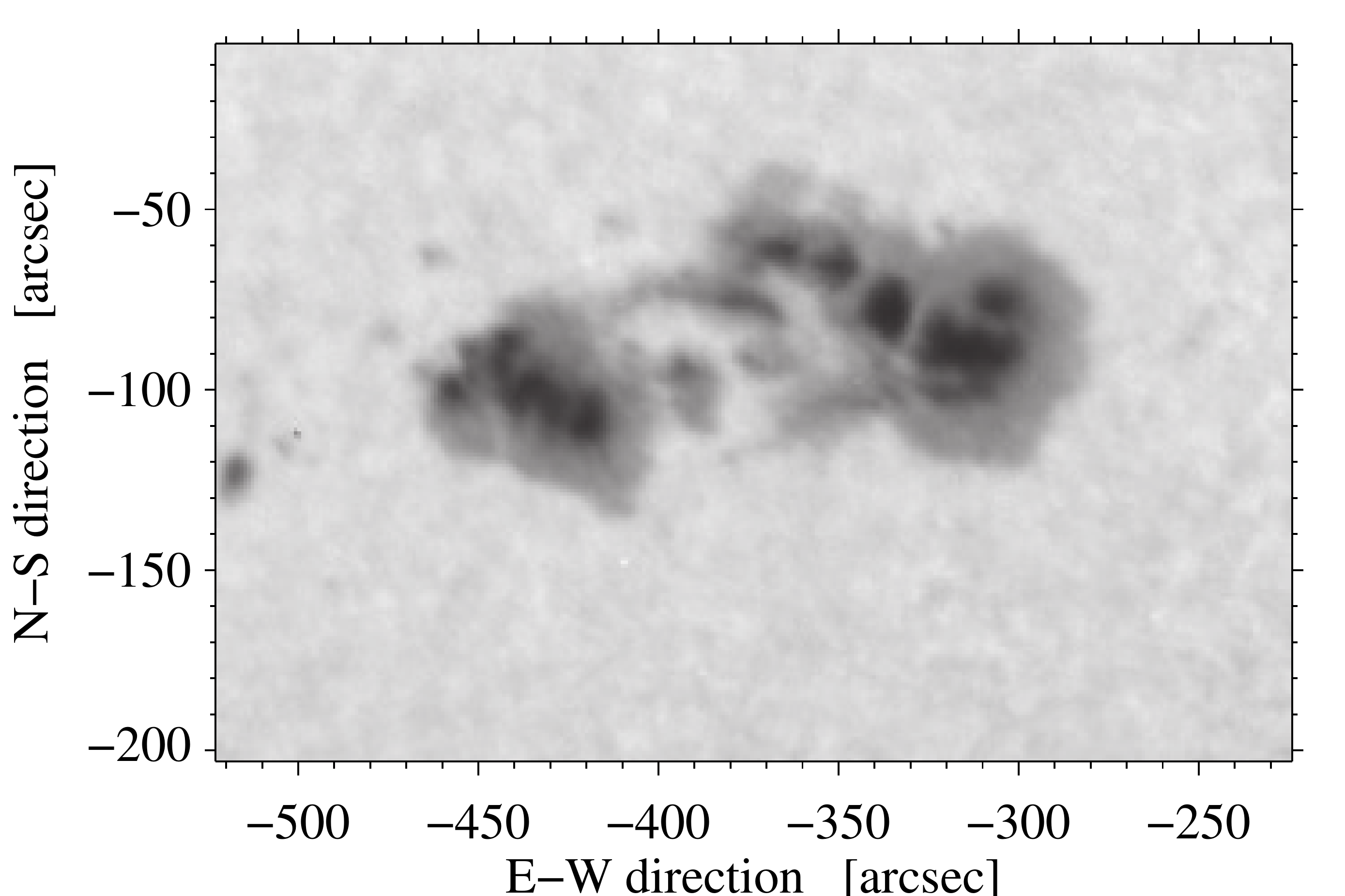}
\includegraphics[width=0.49\textwidth]{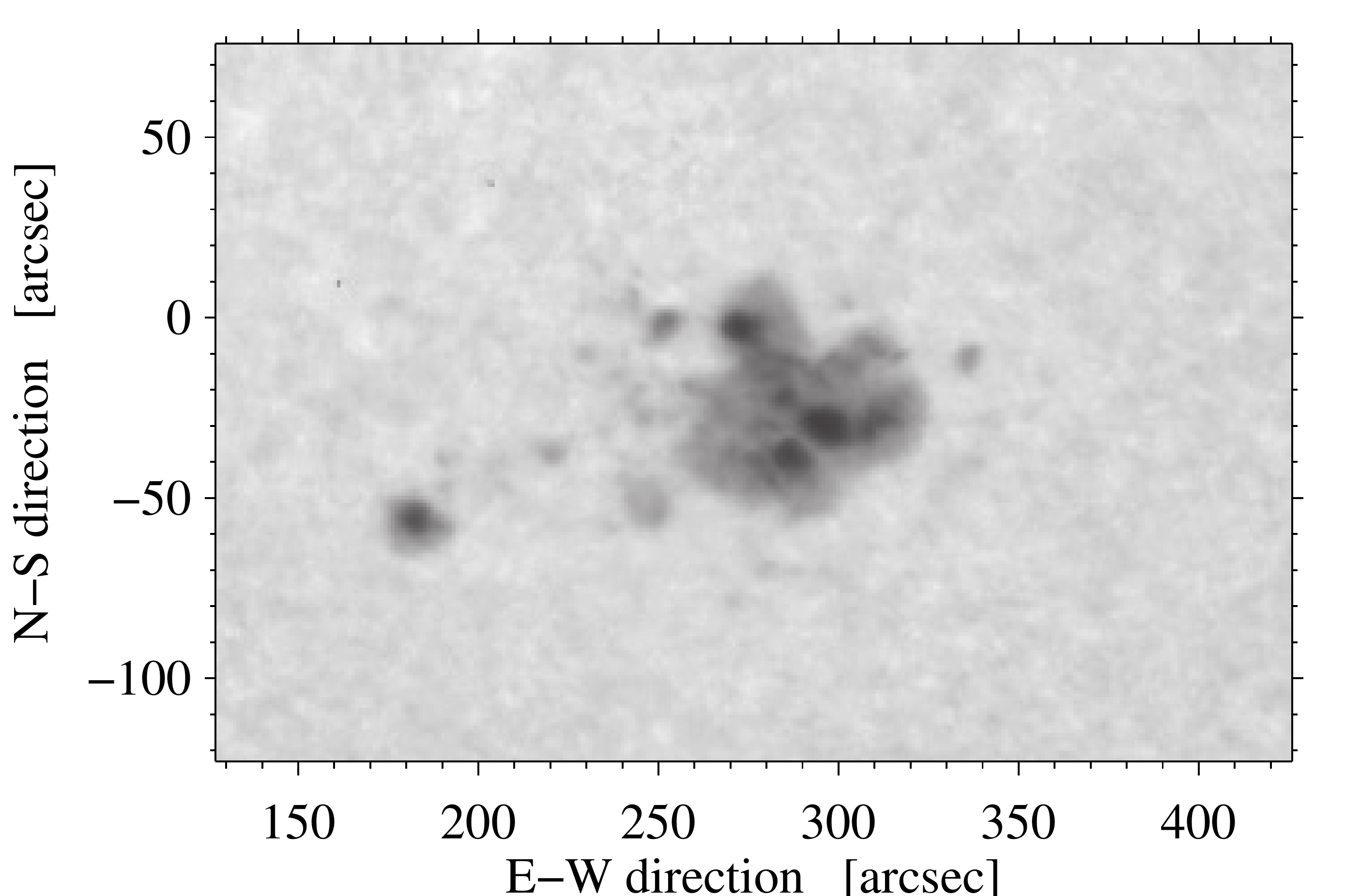}}
\caption{Detailed view of two major sunspot groups observed on 1949 February~3. Sunspot fine structure is clearly discernible after correcting the center-to-limb variation.}
\label{FIG07}
\end{figure*}

Even though the aperture of the Einstein Tower is $D = 60$~cm, it was typically stopped down to $D = 10$ or 15~cm. This delivers more contrast-rich full-disk images considering the prevailing seeing conditions at Potsdam, which can occasionally reach one second of arc or better. In most cases, the seeing conditions are not better than two seconds of arc. Therefore, for time-series data spanning about half a century, spatial resolution can be sacrificed to improve photometric precision. An image scale of 1\arcsec\ pixel$^{-1}$ is a good choice. Thus, the $7086 \times 7086$-pixel images were first enlarged to $8192 \times 8192$-pixel images using cubic spline interpolation before using a $4 \times 4$-pixel boxcar average to raise the photometric precision from 8~bit to about 10~bit for the final $2048 \times 2048$-pixel images (Level~1 and~2 data, see Sect.~\ref{LEVEL}). Since the precision of the calibrated data products is less then 16-bits, normalizing the images offers advantages. The mean value of the quiet-Sun disk-center intensity was set to unity, which conserves the gain in sensitivity. The resulting floating-point images are well-conditioned for storage as scaled 16-bit integer FITS images with optional gzip compression.

Visual and automatic inspection were used to identify photographic data, which are not suitable for further data processing. Criteria for rejection include overexposed images, scan lines caused by jittery motion of the scanning slit camera, occasionally interspersed high-resolution images, cracked photographic plates, bright image background (strong stray light), punch holes in photographic films, double exposures, scratches on photographic plates, artifacts in the emulsion, presence of clouds, and missing markings of the image orientation (see Sect.~\ref{LEVEL1}).

The next critical step concerns determining the solar radius and the disk-center coordinates. An accuracy of one to two pixels is needed, otherwise artifacts will become easily discernible in the limb-darkening corrected images (see below). Two methods were employed, which both lead in most cases (around 95\%) to good results, i.e., they deliver results with the aforementioned accuracy. If the results disagree (around 4\%), the limb-darkening corrected images are visually inspected and the results from the best performing method are chosen. Only in a few cases (around 1\%) of low-quality images, the disk-center coordinates had to be determined manually. The results of this processing step are stored to facilitate automatic reprocessing and future updating of the database.

The solar disk is in general well centered on the plate so that the inner $1024 \times 1024$ pixel can be used to get an estimate of the mean intensity of the solar disk. A threshold of 1/6 of the mean intensity yields a still noisy, binary mask of the disk. Noise is removed with morphological opening using a circular structuring element with a diameter of 32~pixels. The first method simply derives the disk-center coordinates from the center-of-gravity of the mask, and the radius follows from the sum of the masked pixels using the area formula for a circular disk. The second method first identifies the largest region of contiguous pixels in the mask. Thus, all smaller noise structures are eliminated. The image is then multiplied with the cleaned mask before computing two intensity traces by averaging all columns and rows. The intensity profiles along the $x$- and $y$-axis are smoothed with a boxcar average, where the width of the smoothing window is 50~pixels. A second order polynomial is fitted to the central part of the intensity profiles, which yields the disk-center coordinates. A first estimate of the radius is derived from the sum of all pixels, where the counts of the 8-bit image exceeds a threshold of 20, i.e., analogous to the first method. An improved determination of coordinates and radius is obtained with a Sobel edge enhancement operation selecting pixels above a certain threshold. If the distance of these pixels from the estimated disk center deviates by more than twice the respective rms-deviation, the pixels are discarded. A circle is then fitted to the remaining pixels to improve the initial estimate. This procedure is repeated until all pixels fall within twice the rms-deviation limits. The second method implements an improved version of the algorithm that is used for calibrating BBSO full-disk data \citep{Denker1999a}. As a sanity check, ellipse fitting was applied to a random sample of full-disk images. The results yield on average a 2.5\arcsec\-difference between semi-major and -minor axis. The mean value of semi-major and -minor axis matches matches the solar radius determined in circle fitting within 0.1\arcsec. The disk-center coordinates of circle and ellipse fitting agree within 0.1\arcsec. Thus, we did not corrected elliptical distortions. The calibration and centering results are depicted in Fig.~\ref{FIG06}{\ns}b.

\begin{figure*}[t]
\centerline{\includegraphics[width=\textwidth]{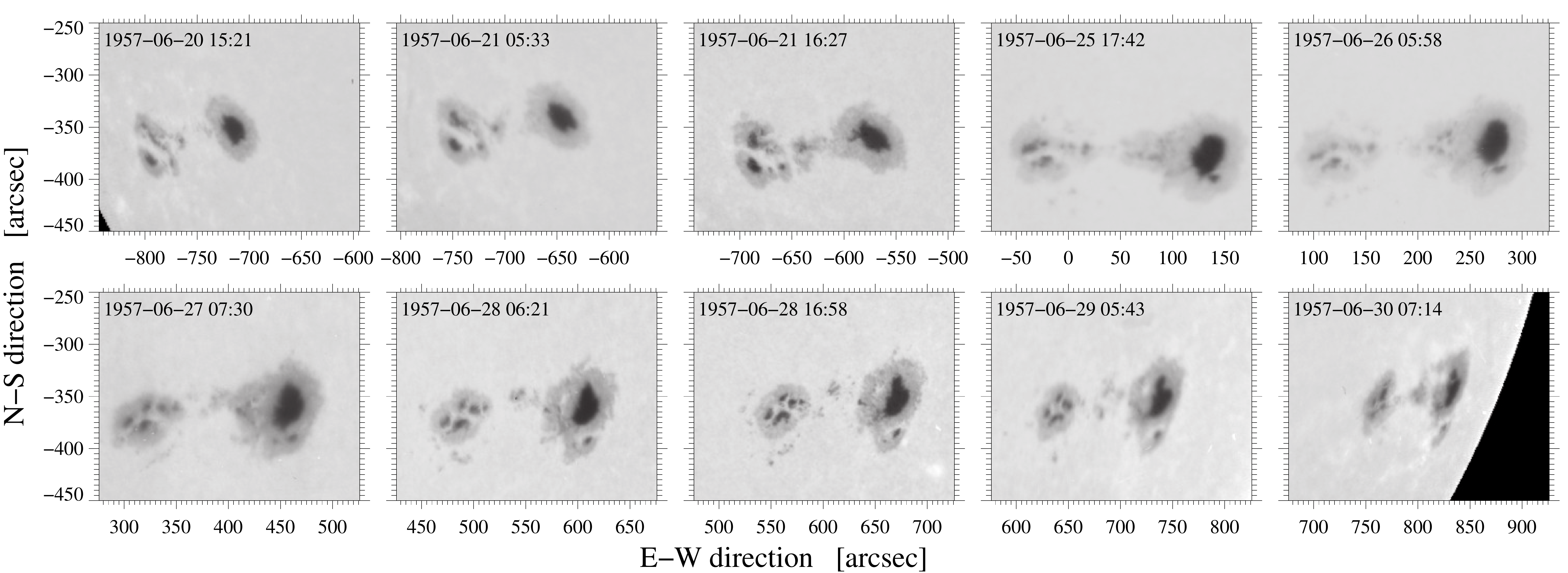}}
\caption{Time-series depicting the disk passage of an active region starting at
the eastern limb on 1957 June~20 and reaching western limb on 1957 June~30.}
\label{FIG08}
\end{figure*}

The computation of the limb darkening profile also follows \citet{Denker1999a}. First, a radial intensity profile is computed by taking an azimuthal average of all disk pixels. The inner 40\% of the profile are fitted by a second order polynomial. Strongly deviating values, which are caused by the presence of dark sunspots and bright plages, are flagged and ignored when the fitting procedure is repeated. The polynomial fit is then merged with the observed radial intensity profile and expanded to a two-dimensional limb-darkening function. A mask based on this function also makes it possible to remove artefacts beyond the solar disk (see Fig.~\ref{FIG06}{\ns}c). In Level~2 data, this two-dimensional limb-darkening function is subtracted from the full-disk image to obtain a contrast enhanced image (see Fig.~\ref{FIG06}{\ns}d), which facilitates better visual inspection, for example, of the sunspot fine structure. Finally, the images are resampled with a fixed image scale of 1\arcsec\ pixel$^{-1}$.

\newpage

The solar full-disk images in Fig.~\ref{FIG06}{\ns}b\,--\,d are already presented in the standard orientation, i.e., with $P$-angle correction and North at the top and East to the left. The rotation angles for all the selected images were derived by calculating the position of west point $q$ according to the equations given by \citet{vonKlueber1948}
\begin{equation}
    q = 270^{\circ} + A - P + \alpha \quad \mathrm{and}    
\end{equation}
\begin{equation}
    \sin P = \frac{\cos \varphi}{\cos \delta}\sin A,\rule[-4mm]{0mm}{4mm}
\end{equation}
where $\varphi$ is the pole-height, $\delta$ the declination of the Sun, $p$ the solar $P$-angle, $A$ the coelostat azimuth, and $\alpha$ the angle of the $45^{\circ}$ turning mirror in the optical laboratory. This formula gives the west point angle with respect to an orientation, where North is at the bottom and West to the left. All images showed an offset of $-90^\circ$ for west point when using these two equations, which had to be corrected for the World Coordinate System (WCS). The value for $A$ as well as the observing date and times were obtained from the observer's log. The coelostat azimuth values in the log were in the range of $0^\circ$ to $180^\circ$, which had to be converted to the range $-90^\circ$ to $+90^\circ$. The values for $\delta$ and $P$ were queried from the JPL HORIZONS on-line solar system data and ephemeris computation service.\footnote{\href{https://ssd.jpl.nasa.gov/?horizons}{ssd.jpl.nasa.gov/?horizons}} The ephemerides were computed for the ICRF/J2000.0 reference frame for geometric and astrometric quantities. Including WCS information is a crucial step in conditioning the Einstein Tower full-disk images for scientific use. Although, most plates contained the required information for this step, a sanity check by visual inspection is performed. For plates with missing information the required values from proceeding and succeeding plates are used. Figures~\ref{FIG07}{\ns} and~\ref{FIG08}{\ns} depict examples, of sunspot groups and active regions, which can be used for studying the fine structure and temporal evolution of sunspots.


\section{Data Levels}\label{LEVEL}

The scanned full-disk data were saved as 8-bit TIFF images with $7086 \times 7086$ pixels. The entire scanned dataset comprises 3620 photographic negative images. These data are the point of origin for the three data levels described below. The file naming convention\smallskip\\ \texttt{soet\_white\_f[t]\_[yyyymmddd]\_[hhmmss][c].fts},\smallskip\\ where the atoms of the filename correspond to the observatory (\texttt{soet} $\rightarrow$ solar observatory Einstein Tower), the observed wavelength range (\texttt{white} $\rightarrow$ broad-band white-light image), a two-letter flag describing the image type (\texttt{f} $\rightarrow$ full-disk image, \texttt{t} = \texttt{i} $\rightarrow$ raw image, \texttt{t} = \texttt{l} $\rightarrow$ calibrated image, and \texttt{t} = \texttt{r} $\rightarrow$ limb-darkening corrected image), 8-character observing date (\texttt{yyyymmdd} $\rightarrow$ year | month | day), and 6-character observing time (\texttt{hhmmss} $\rightarrow$ hour | minute | second). An extra character (\texttt{c} $\rightarrow$ \texttt{a}, \texttt{b}, \texttt{c}, \ldots) is added at the end of the filename to distinguish the images, which have the same observed date and time. In cases, where no time information was given, the most likely time (see Figure~\ref{FIG03}\ns), i.e., 08:00~UT was inserted. In some cases, when multiple images were observed on a given date, not all images had the time information available. Under the assumption that the plates were recorded one after the other, a reasonable time difference of 10~min was adopted for images without time information. The overall file naming approach follows the conventions used for BBSO\footnote{\href{https://www.bbso.njit.edu/Archive/filenames.html}{www.bbso.njit.edu/Archive/filenames.html}} and SoHO data and facilitates to retrieve basic information just from the filenames. Thus, further processing in data pipelines becomes possible because the order of the atoms in the file names makes sorting data straightforward.

Objects in APPLAUSE are sorted according to their types as plates (as in the case of the Einstein Tower full-disk images), log books, notes and envelopes. Each Cultural Heritage Object (CHO) has a unique APPLAUSE Identifier (AID) stored in corresponding table column of the APPLAUSE database. A Digital Object Identifier (DOI) can be constructed in the following way\smallskip\\ \texttt{doi:10.1876/plate/[AID]},\smallskip\\
where the AID/DOI is formed by\smallskip\\
\texttt{AIP-Prefix/plate/[DR]/[archive\_id]\_[seq\_number]}.\smallskip\\
The DOI prefix for AIP is \texttt{AIP-Prefix} = \texttt{10.1876}, and the solar plates will be published as part of Data Release 3 (DR3) with the version number \texttt{DR} = \texttt{dr.3s}. The Einstein Tower plates are a new archive in the APPLAUSE collection, which can be uniquely identified by the \texttt{archive\_id} = \texttt{006}. The leading zero indicate that the solar full disk data are the sixth archive, which originates from AIP. Other observatories use higher archive identifiers, e.g., Bamberg archives start with \texttt{archive\_id} = \texttt{201}. The three data levels below, which were derived from one digitized full-disk image, are considered as one CHO. Since these data are easily sorted in chronological order, the first plate observed at 08:00~UT on 1943 April 19 receives the \texttt{seq\_number} = \texttt{1}, which initializes the whole sequence. The corresponding DOI is\smallskip\\
\texttt{doi:10.1876/plate/dr.3s/006\_0001}.


\subsection{Level~0 -- Raw data}

Conversion from TIFF to FITS and from photographic negatives to positives made it possible to include more descriptive headers and to ease visualisation. No information is lost, and the process is fully reversible. Thus, Level~0 data can be used by other researchers in other data processing pipelines and also ensures that the full-disk data can be reprocessed once improved data processing techniques become available. The original photographic data is still stored at the Einstein Tower and will be preserved as an important historical data record. Full-disk images of all data levels are stored in FITS format \citep{Wells1981, Hanisch2001}. The FITS header of the raw Level~0 data is summarized in Table~\ref{TAB01}{\ns} using the data displayed in Fig.~\ref{FIG06}{\ns} as an example. The keywords \texttt{SQUALITY} and \texttt{PQUALITY} refer to the seeing and photographic quality, respectively. These values are taken from the observer's log book. In both cases, a scale of 1\,--\,4 (bad, mediocre, good, and very good) is used, which sometimes uses ranges or qualifiers to indicate a finer scale. Note that the Einstein Tower observer's log books are not yet digitized. The FITS history and comment fields are given in Table~\ref{TAB02}{\ns}, which contains the history for all three date levels.


\subsection{Level~1 -- Calibrated data}\label{LEVEL1}

Out of 3620 files that were analyzed, 3571 images were identified for processing. Automatic radius and center estimates were derived for 3446 images while 125 images required some tweaking in the algorithm. In the end, only 49 images were discarded because they were unsuitable for further processing. However, they are still available as Level~0 data. Level~1 data are characterized by resampling from $7086 \times 7086$ pixel to $2048 \times 2048$ pixels, enhancing photometric precision. The solar disk was centered, and disk-center center coordinates as well as the radius of the solar disk were determined. The precise image scale was derived and fixed to 1\arcsec\ pixel$^{-1}$. A two-dimensional limb-darkening function was calculated for each full-disk image. However, at this data level it was only used to eliminate artefacts outside the solar disk. At this data level, the full-disk images have the standard solar orientation so that they are ready for scientific use. Updated and new entries in the FITS header are summarized in Table~\ref{TAB03}{\ns} for calibrated Level~1 data. In particular, date \& time information and helioprojective-cartesian coordinates according to the WCS standards were added to facilitate data analysis in the same way as for contemporary data.

\begin{table}[t]
\fontsize{6pt}{6.8pt}\selectfont
\rule{\columnwidth}{1.2pt}\vspace*{-6pt}
\rule{\columnwidth}{0.4pt}
\textbf{Keyword} \hspace{6pt} \textbf{Value} \hspace{54pt} \textbf{Comment}\vspace*{-3pt}\\
\rule{\columnwidth}{0.4pt}\vspace*{-5pt}
\begin{verbatim}
SIMPLE  =                    T /                                                 
BITPIX  =                    8 / number of bits per data pixel                   
NAXIS   =                    2 / number of data axes                             
NAXIS1  =                 7086 /                                                 
NAXIS2  =                 7086 /                                                 
ORIGIN  = 'Leibniz Institute for Astrophysics Potsdam (AIP)' / FITS file origin  
TIMESYS = 'UTC     '           /                                                 
DATE    = '2020-02-04T15:18:30' / date of last modification                      
DATE_OBS= '1949-02-03T13:36:00.000' / recorded date of the observation           
FILENAME= 'soet_white_fi_19490203_133600.fts' / filename of this file           
FN-LOGB =                    T / log book entry is available                     
OBJECT  = 'Sun     '           / object                                          
OBJTYPE = 'full-disk'          / object type                                     
OBSERVAT= 'Solar Observatory Einstein Tower' / observatory name                  
SITENAME= 'Telegraphenberg-Potsdam, Germany' / observatory site name                           
TELESCOP= 'Einstein Tower'     / telescope name                                  
SITELONG=            13.063855 / east longitude of the observatory [deg]         
SITELAT =            52.378750 / latitude of the observatory [deg]               
SITEELEV=                87.00 / elevation of the observatory [m]                
TELAPER =             0.600000 / clear aperture of the telescope [m]             
TELFOC  =              14.0000 / focal length of the telescope [m]               
TELSCALE=              14.7300 / plate scale of the telescope [arcsec/mm]        
AZIMUTH =                  0 / telescope coelostat azimuth [deg]               
REFERENC= 'doi:10.1002/asna.201612442' / reference Einstein Tower                
DOI     = '10.17876/plate/dr.3s/006_0767' / DOI of the plate                
LICENSE = 'https://creativecommons.org/licenses/by/4.0/' /                 
INSTRUME= 'Scanning Slit Camera' / instrument                                    
DETNAM  = 'photographic plate' / detector                                        
METHOD  = 'direct photograph'  / method of observation                           
FILTER  = 'Schott Glas'        / filter type
WAVELNTH=                 4200 / central wavelength [Angstrom]                   
WAVEMIN =                 3850 / minimum wavelength [Angstrom]                   
WAVEMAX =                 4700 / maximum wavelength [Angstrom]                   
OBS_TYPE= 'BBAND   '           / Schott Glass GG3 and BG12                      
EXPTIME =           0.00250000 / exposure time [s]                               
NUMEXP  =                    1 / number of exposures on the plate                
SQUALITY= '2        '          / seeing quality
PLATEFMT= '15x15   '           / plate format in cm                              
PLATESZ1=              15.0000 / plate size along axis 1 [cm]                    
PLATESZ2=              15.0000 / plate size along axis 2 [cm]                    
PLATENUM= '698_1   '           / plate number in the observation catalogue
PQUALITY= '2       '           / quality of the plate
SCANDATE= '2009-07-03T18:47:58' / date of plate digitization 
SCANNER = 'Epson Perfection 4990' / scanner model name                           
SCANRES1=                 1200 / scan resolution along axis 1 [dpi]              
SCANRES2=                 1200 / scan resolution along axis 2 [dpi]             
PIXSIZE1=              21.1685 / pixel size along axis 1 [microns]               
PIXSIZE2=              21.1685 / pixel size along axis 2 [microns]               
SCANSOFT= 'Epson Scan'         / name of the scanning software                   
SCANAUTH= 'J. Rendtel, C. Denker, P.S. Pal' / author of scan                    
VERSION = 'v1.0    '           / data processing version                         
DATA-LEV=                    0 / raw data                                        
FLAG    =                    1 / [1] image, [2] time, [4] azimuth exist          
XSCALE  =            0.0211685 / image scale (X) [mm/pixel]                      
YSCALE  =            0.0211685 / image scale (y) [mm/pixel]                      
CTYPE1  = 'x-direction'        /                                                 
CRPIX1  =                    0 / reference point x [pixel]                       
CDELT1  =            0.0211685 / increment x [mm]                                
CUNIT1  = 'mm      '           /                                                 
CTYPE2  = 'y-direction'        /                                                 
CRPIX2  =                    0 / reference point y [pixel]                       
CDELT2  =            0.0211685 / increment y [mm]                                
CUNIT2  = 'mm      '           /                                                 
DEC     =             -15.6976 / declination of the Sun [deg]                   
RA      =              21.1734 /rectascension of the Sun [deg]                  
ELEV    =              20.8087 / elevation of the Sun (disk center) [deg]       
AIR-MASS=              3.74700 / airmass                                        
SOLAR-AZ=              7.64332 / azimuth of the Sun  [deg]                      
SOLAR-P0=             -13.1781 / solar axis position angle [deg]                
SOLAR-B0=             -6.23452 / solar axis tilt angle [deg]                    
SOLAR-L0=              259.931 / Carrington longitude [deg]                     
SOLAR-R =              975.037 / Solar radius [arcsec]                          
CAR_ROT =              1276.28 / Carrington rotation number                     
SID-TIME=              324.340 / siderial time [deg]                            
DATAVALS=             50211396 / number of pixels                               
DATAMIN =                    0 / minimum value                                  
DATAMAX =                  255 / maximum value                                  
DATAMEDN=              99.0000 / median value                                   
DATAMEAN=              96.2548 / mean value                                     
DATARMS =              80.9244 / rms deviation                                  
DATASKEW=             0.140546 / skewness from the mean value                   
DATAKURT=             -1.57882 / kurtosis
DATASUM = '1461316397'         / data unit checksum created                     
CHECKSUM= 'oA64r063o763o763'   / HDU checksum created  
\end{verbatim}\vspace*{-10pt}
\rule{\columnwidth}{0.4pt}\vspace*{-5pt}
\rule{\columnwidth}{1.2pt}
\caption{Primary FITS header with keywords, values, and comments for Level~0 raw data based on the scanned TIFF images.}
\label{TAB01}
\end{table}


\subsection{Level~2 -- Limb-darkening corrected data}

Level~2 data are currently the final products of the data reduction. Subtraction of the previously determined two-dimensional limb-darkening function yields contrast enhanced full-disk images. Since the characteristic curve of the photographic material is unknown and gray level scales are only present for select observation, further photometric calibration is hampered. However, the limb-darkening corrected images can be multiplied with any theoretical two-dimensional limb-darkening function as a first-order approximation of the intensity distribution across the solar disk. Updated and new entries in the FITS header are summarized in Table~\ref{TAB04}{\ns} for limb-darkening corrected Level~2 data. In summary, all 3620 plates are available as Level~0 images and 3571 are part of Level~1 and 2 data.

\begin{table}[t]
\fontsize{6pt}{7pt}\selectfont
\rule{\columnwidth}{1.2pt}\vspace*{-6pt}
\rule{\columnwidth}{0.4pt}
\textbf{Keyword} \hspace{1pt} \textbf{History \& Comment}\vspace*{-3pt}\\
\rule{\columnwidth}{0.4pt}\vspace*{-5pt}
\begin{verbatim}
HISTORY
HISTORY negative to positive transformation (TIFF to FITS conversion)            
HISTORY FITS header added with metadata
\end{verbatim}\vspace*{-12pt}
\hspace*{-1pt}\rule{\columnwidth}{0.4pt}\vspace*{-8pt}
\begin{verbatim}
HISTORY center image and P-angle correction
HISTORY scale image to 1 arcsec/pixel    
HISTORY update FITS header and add WCS information
HISTORY remove artifacts outside solar disk
\end{verbatim}\vspace*{-12pt}
\hspace*{-1pt}\rule{\columnwidth}{0.4pt}\vspace*{-8pt}
\begin{verbatim}
HISTORY subtract two-dimensional limb-darkening function
\end{verbatim}\vspace*{-12pt}
\hspace*{-1pt}\rule{\columnwidth}{0.4pt}\vspace*{-8pt}
\begin{verbatim}
HISTORY
COMMENT                                                                          
COMMENT Reference publications:                                                  
COMMENT                                                                          
COMMENT - von Klueber, H. &  Mueller, H. 1948: Bemerkungen zur Technik 
COMMENT   direkter Sonnenaufnahmen mit langbrennweitigen Instrumenten. 
COMMENT   Z. Astrophys. 24, 207                                                                   
COMMENT - Denker, C., Heibel, C., Rendtel, J., et al. 2016: Solar physics
COMMENT   at the Einstein Tower. Astron. Nachr. 337, 1105                              
COMMENT - Pal, P.S., Denker, C., Rendtel, J., et al. 2020: Solar      
COMMENT   observatory Einstein Tower - Data release of the digitized solar       
COMMENT   full-disk photographic plate archive. Astron. Nachr., submitted                  
COMMENT                                                                          
COMMENT Publications based on this digitized photographic plate 
COMMENT are requested to include the following acknowledgement.                                  
COMMENT                                                                          
COMMENT Based on photographic data obtained at the Einstein Tower      
COMMENT of the Leibniz Institute for Astrophysics Potsdam (AIP).
COMMENT
\end{verbatim}\vspace*{-10pt}
\rule{\columnwidth}{0.4pt}\vspace*{-5pt}
\rule{\columnwidth}{1.2pt}
\caption{History and comment fields. The horizontal lines indicate the Level~0\,--\,2 data processing steps.}
\label{TAB02}
\end{table}


\section{Discussions and conclusions}

This work introduced photographic full-disk images obtained at the solar observatory Einstein Tower. Digitization, storage in a database, and incorporation of meaningful metadata are only the first steps in scientifically exploiting this now broadly available resource. The image quality is in general high and the coverage is fairly good in the first 25 years of observations, which makes it possible to track solar activity. Even when the full-disk data are sparse, they complement similar observations from around the world. The next steps in data processing will include feature recognition, which will provide size, location, and photometry of active regions, sunspots (including umbrae and penumbrae), and pores. The spatial resolution and dynamic range of the contrast enhanced images is also sufficient to identify facular regions near the limb. Applications include the determination of sunspot group tilt angles \citep[\textit{e.g.}][]{SenthamizhPavai2016}, penumbra/umbra ratio \citep[\textit{e.g.}][]{Brandt1990, Tlatov2019}, umbral size and minimum intensity \citep[\textit{e.g.}][]{Kiess2014}, maximum magnetic field and umbral/penumbral areas \citep[\textit{e.g.}][]{Watson2011}, and the statistics of pores \citep[\textit{e.g.}][]{Verma2014}, among others.

\begin{table}[t]
\fontsize{6pt}{7pt}\selectfont
\rule{\columnwidth}{1.2pt}\vspace*{-6pt}
\rule{\columnwidth}{0.4pt}
\textbf{Keyword} \hspace{6pt} \textbf{Value} \hspace{54pt} \textbf{Comment}\vspace*{-3pt}\\
\rule{\columnwidth}{0.4pt}\vspace*{-5pt}
\begin{alltt}
SIMPLE  =                    T /                                                
BITPIX  =                   16 / number of bits per data pixel                  
NAXIS   =                    2 / number of data axes                            
NAXIS1  =                 2048 / number of pixels in E-W direction              
NAXIS2  =                 2048 / number of pixels in N-S direction              
BSCALE  =              1.00000 /                                                
BZERO   =              0.00000 /                                                
EXTEND  =                    T /  FITS dataset may contain extensions
\vdots  \hspace*{1cm} \vdots     \hspace*{1cm}   \vdots
DATA-LEV=                    1 / calibrated data
THETA   =             -153.826 / rotation angle [deg]                           
FLIP    =                    0 / flip [2] 180 deg, [5] horizontal, [7] vertical 
CTYPE1  = 'HPLN-TAN'           / North (top) / South (bottom)                   
CRPIX1  =                 1024 / reference point x [pixel]                      
CDELT1  =              1.00000 / coordinate increment x [arcsec]                
CUNIT1  = 'arcsec  '           /                                                
CTYPE2  = 'HPLN-TAN'           / East (left) / West(right)                      
CRPIX2  =                 1024 / reference point y [pixel]                      
CDELT2  =              1.00000 / coordinate increment y [arcsec]                
CUNIT2  = 'arcsec  '           /                                                
XSCALE  =              1.00000 / image scale (X) [arcsec/pixel]                 
YSCALE  =              1.00000 / image scale (y) [arcsec/pixel]                 
XCEN    =                 1024 / disk center (X) [pixel]                        
YCEN    =                 1024 / disk center (Y) [pixel]
\vdots  \hspace*{1cm} \vdots     \hspace*{1cm}   \vdots
DATAVALS=              2990878 / number of on-disk pixels                       
DATAMIN =           0.00362179 / minimum value                                  
DATAMAX =              254.523 / maximum value                                  
DATAMEDN=              160.479 / median value                                   
DATAMEAN=              153.624 / mean value                                     
DATARMS =              49.4269 / rms deviation                                  
DATASKEW=            -0.428141 / skewness from the mean value                   
DATAKURT=            -0.726274 / kurtosis                                       
DATASUM = '2084252676'         / data unit checksum created 2019-08-14T20:10:51 
CHECKSUM= 'RdgRSZePRbePRZeP'   / HDU checksum created 2019-08-14T20:10:51       
OBSCALE=     0.00388370733708 / Original BSCALE Value                          
OBZERO =        127.261322021 / Original BZERO Value
\end{alltt}\vspace*{-10pt}
\rule{\columnwidth}{0.4pt}\vspace*{-5pt}
\rule{\columnwidth}{1.2pt}
\caption{Updated and new FITS header keywords, values, and comments for calibrated Level~1 data.}
\label{TAB03}
\end{table}

In the era of machine learning, its application to the sunspot detection was already explored by many. \citet{Goel2014} described the automated detection of sunspots in full-disk continuum images obtained with the Michelson Doppler Imager \citep[MDI,][]{Scherrer1995} on board the Solar and Heliospheric Observatory \citep[SoHO,][]{Domingo1995}. Going one step further, \citet{Yang2018} employed artificial intelligence to automatically detect sunspots in continuum images obtained with the Helioseismic and Magnetic Imager \citep[HMI,][]{Scherrer2012} on board the Solar Dynamics Observatory \citep[SDO,][]{Pesnell2012}. The Einstein Tower full-disk data are well-conditioned to employ similar techniques for the automatic sunspot detection.

Astronomical observatories across Germany have substantial collections of photographic plates. \citet{Tsvetkova2009} presented a digitized database of the Carte du Ciel plates stored at AIP. Following their work, \citet{Groote2014} described the cooperation of the Potsdam and Bamberg observatories in digitizing historic photographic plates. The combined database comprises about 80,000 astronomical photographic plates, which lays the foundation for the APPLAUSE project, where the digitized plates are made accessible by virtual observatory tools. The APPLAUSE project aims to digitize photographic plates available in the archives of the Hamburg, Bamberg, and Potsdam observatories and to make them available via a dedicated database. The archives do not only include the photographic plates of stellar observations and surveys but also log books, envelopes and observer notes, to preserve also the cultural heritage within the digital archive. The solar observatory Einstein Tower was the first solar tower telescope in Europe and one of the most prominent places for solar research until World War~II. The observatory collected photographic plates during the period 1943\,--\,1990. As part of the APPLAUSE project, the digitization of the photographic plates with solar full-disk images was described in this work, including the various steps from initial data processing to final products. This reference publication is intended to make the full-disk data of the Einstein Tower available for scientific use, which will be publicly accessible for interested scientists and the general public.

\begin{table}[t]
\fontsize{6pt}{7pt}\selectfont
\rule{\columnwidth}{1.2pt}\vspace*{-6pt}
\rule{\columnwidth}{0.4pt}
\textbf{Keyword} \hspace{6pt} \textbf{Value} \hspace{54pt} \textbf{Comment}\vspace*{-3pt}\\
\rule{\columnwidth}{0.4pt}\vspace*{-5pt}
\begin{alltt}
SIMPLE  =                    T / file does conform to FITS standard\rule{0mm}{3.mm}
\vdots  \hspace*{1cm} \vdots     \hspace*{1cm}   \vdots
DATA-LEV=                    2 / limb darkening corrected data 
\vdots  \hspace*{1cm} \vdots     \hspace*{1cm}   \vdots
DATAVALS=              4194304 / number of on-disk pixels                       
DATAMIN =    3.61910501567E-12 / minimum value                                  
DATAMAX =        3.75712860587 / maximum value                                  
DATAMEDN=       0.957298137804 / median value                                   
DATAMEAN=       0.717590655467 / mean value                                     
DATARMS =       0.461805607359 / rms deviation                                  
DATASKEW=      -0.857423473890 / skewness from the mean value                   
DATAKURT=       -1.14990102995 / kurtosis                                       
DATASUM = '3901920496'         / checksum created 2020-02-10T16:48:05 
CHECKSUM= '0iFa0g9S0gEY0g9Y'   / checksum created 2020-02-10T16:48:05  
\end{alltt}\vspace*{-10pt}
\rule{\columnwidth}{0.4pt}\vspace*{-5pt}
\rule{\columnwidth}{1.2pt}
\caption{Updated and new FITS header keywords, values, and comments for limb-darkening corrected Level~2 data.}
\label{TAB04}
\end{table}

A physical preservation of all the photographic plates and connected material like log books is very problematic and costly, because of the decay of photographic emulsions or bad paper quality. A digital preservation of the contained information is efficient and enables another long-term preservation strategy. As an example, with modern software, programming methods, and available stellar catalogs like Gaia, the immense amount of stellar objects on the plates in APPLAUSE can be identified and are photometrically calibrated to reach modern quality expectations. We gain thus a century-long timeline for stars, their variation and also means to retrospectively identify events, e.g., blazars. Similarly, the digitized historical solar plates complement contemporary synoptic data and various data products. For example, historical digitized data can be adapted to existing frameworks developed to provide easy access to current data, e.g., the Heliophysics Event Knowledgebase \citep[HEK,][]{Hurlburt2012}.

For accommodating the requirements of metadata systems, the APPLAUSE project suggested an extension of the FITS header for digitised astronomical photographic plates. All published digitised items in FITS format from the APPLAUSE carry this header. The data tables carry also metadata, reflecting FITS header entries and additional information, which meet most requirements of \textit{Findable, Accessible, Interoperable, and Reusable} (FAIR). For the solar photographic plates, an additional subset of FITS header entries was discussed above. The data tables pertaining to the solar photographic plates will on the one hand help to quickly find, select and retrieve data, and later provide physical parameters characterising located events on the individual plates.


\section*{Acknowledgments}

Funding for APPLAUSE was provided by Deutsche Forschungsgemeinschaft (DFG), Leibniz Institute for Astrophysics Potsdam (AIP), Dr.\ Remeis Sternwarte Bamberg (Friedrich-Alexander-Universit\"at Erlangen-N\"urnberg), Hamburger Sternwarte, and Tartu Observatory. This study was supported by grants DE~787/5-1 of the DFG and 18-08097J of the Czech Science Foundation. In addition, the support by the European Commission's Horizon 2020 Program under grant agreements 824064 (ESCAPE -- European Science Cluster of Astronomy \& Particle physics ESFRI research infrastructures) and 824135 (SOLARNET -- Integrating High Resolution Solar Physics) is highly appreciated. PSP likes to thank the Bhaskaracharya College of Applied Sciences, University of Delhi, for the generous support of his stay as a visiting scientist at AIP. SJGM  acknowledges  the  support  of  the  project  VEGA 2/0048/20. This research has made use of NASA's Astrophysics Data System. We strongly appreciate the efforts by Tobias Benthin, Dennis Collatz, Florian Fischer, Franka Hahn, Celina Liebenow, Philipp Mahncke, Tina Marie Patzner, Thomas Schmidt, Daniel Simanowitsch, and Jannick Werner in digitizing the solar plates as part of their high-school student internship at AIP. We thank Drs.\ Rainer Arlt and Axel Hofmann for their comments and helpful suggestions, which significantly improved the manuscript. We are grateful to the referee Dr. Alexei Pevtsov for his constructive input, which benefited the manuscript substantially.


\end{document}